\documentclass[seceq]{ptptex}
\usepackage{graphics}
\newcommand{\beq}{\begin{equation}}
\newcommand{\eeq}{\end{equation}}

\pubinfo{Vol. 127, No. 1, January 2012}
\title{Quantum Eraser Using A Modified Stern-Gerlach Setup}

\author{Tabish \textsc{Qureshi}\footnote{E-mail: tabish@ctp-jamia.res.in} and Zini \textsc{Rahman}\footnote{E-mail: zini@ctp-jamia.res.in}}
\inst{Centre for Theoretical Physics\\ Jamia Millia Islamia,
New Delhi-110025, India.}

\abst{
We propose a new setup to demonstrate quantum eraser, using
spin-1/2 particles in a modified Stern-Gerlach setup, with a
double slit. The {\em which-way} information can be erased simply by
applying a transverse magnetic field with an additional magnet, resulting in the
emergence of two complementary staggered interference patterns.
Use of the classic Stern-Gerlach setup, and the unweaving of the washed
out interference without any coincident counting, is what makes this
proposal novel. 
}

\begin{document}
\maketitle

\section{Complementarity and Quantum Eraser}
It is well known that particles and light both, are capable of exhibiting a
dual nature. This is commonly referred to as wave-particle duality. What is
commonly not emphasized, is the fact that these natures are mutually
exclusive - for example, light can act either as a particle, or as a wave at
a time. This has its foundation
in Bohr's complementarity principle \cite{bohr}. It can be best understood in
the context of Young's double slit experiment with particles. Complementarity
principle implies that in such an experiment,
there is a fundamental incompatibility between the {\em ``Welcher-Weg"}, or
which-way information and the observation of interference pattern. Thus
any attempt to obtain information about which slit the particle went through,
necessarily destroys the interference pattern. Replying to Einstein's 
famous thought experiment regarding a recoiling double-slit, Bohr had
demonstrated that the uncertainty in the initial position of the double-slit
is precisely enough to wash out the interference pattern.

However, it turns out that it was just fortuitous that the uncertainty
principle seemed to wash out the interference pattern. It has been argued
that one could have the which-way information without appreciably affecting
the spatial part of the wave function of the particle \cite{druhl}. This
can be done by entanglement of the particle with a variable, playing the
role of a which-way marker. So, uncertainty principle is not the fundamental
reason for washing out of interference in a double-slit experiment -
entanglement is.

The double-slit experiment, with entanglement can be understood in the following
way. Let us now assume that the initial state of the particle was
entangled with a certain degree of freedom so that the state can be written
as:
\begin{equation}
|\psi(r)\rangle = {1\over\sqrt{2}}[|\psi_1(r)\rangle|1\rangle +
|\psi_2(r)\rangle|2\rangle],
\end{equation}
where $|1\rangle$ and $|2\rangle$ are certain normalized and orthogonal states,
and $|\psi_1\rangle$ and  $|\psi_2\rangle$ represent possibilities of the
particle going through one or the other slit.
It is easy to see that when one calculates probability distribution of the
particle on the screen $|\psi(r)|^2$, the cross-terms,
$\psi_1^*(r)\psi_2(r)$ and $\psi_2^*(r)\psi_1(r)$,
which are responsible for interference,
are killed by the orthogonality of $|1\rangle$ and $|2\rangle$.

An interesting idea was put forward by Jaynes \cite{jaynes}, and later
independently by Scully and Dr\"{u}hl \cite{druhl} saying that if the
which-way information is stored in quantum detectors, it could also be
erased by ``reading out" those specific observables of the quantum detectors
which do not distinguish between the two paths. In this situation,
it should be possible to get back the interference. This came to be known
as the {\em quantum eraser} \cite{druhl,jaynes}.
Scully, Englert and Walther proposed an experiment with Rydberg atoms, with
micro-maser cavity detectors acting as which-way markers. They argued that
if one were to perform a correlated measurement of the two detectors in such
a way that the which-way information is lost, the interference pattern
will be visible again \cite{scully}. 

Quantum eraser has been experimentally realized by various people using photons
\cite{mandel,chiao, zeilinger,kim-shih,walborn,kim,andersen}, mainly
because it is easy to produce entangled photons via spontaneous parametric
down conversion (SPDC). There have been some other proposals regarding
NMR analogue of quantum eraser \cite{cory},  neutral kaons \cite{bramoni}
and cavity QED \cite{gsa}. There is also one proposal using atoms in
an optical Strern-Gerlach model \cite{chianello}. A similar argument has
been put forward in the context of Rabi oscillations in atoms  \cite{tuminello}.
Quantum eraser has been demonstrated using atom interferometry
\cite{durr, bertet}.  A quantum  eraser has also been used to generate
entangled coherent states \cite{gerry}.  Effect of decoherence on a
quantum eraser has also been studied \cite{lima}. Recently, manifestation
of geometric phase in the setup for quantum eraser, using polarization states
of light, has been experimentally demonstrated \cite{kitano}.
Polarization states of light have also been used to demonstrate
quantum eraser in some other works \cite{schwindt, schneider}.

Most of these experiments require state of the art technology, either using
entangled photon/particle sources or atom interferometers. Simple
implementations have been demonstrated with photons, but not with particles.
In all such experiments, the ``first-order" interfrence is lost, because
the particles carry which-way information. However, if one does a 
coincident counting of particles with certain specific states of the
which-way marker, the interference pattern is observed.

\section{Quantum Eraser with a modified Stern-Gerlach apparatus}

Here we propose an implementation of quantum eraser using a modified 
Stern-Gerlach setup with spin-1/2 particles. As one will see later,
the nice feature of this implementation is that it is simple, easy to visualize
and carry out, and quantum erasure is vividly brought out by two
sets of interference fringes on a real screen. The setup is fairly simple,
and no coincident counting of any kind is required.

\begin{figure}
\centerline{\resizebox{13cm}{!}{\includegraphics{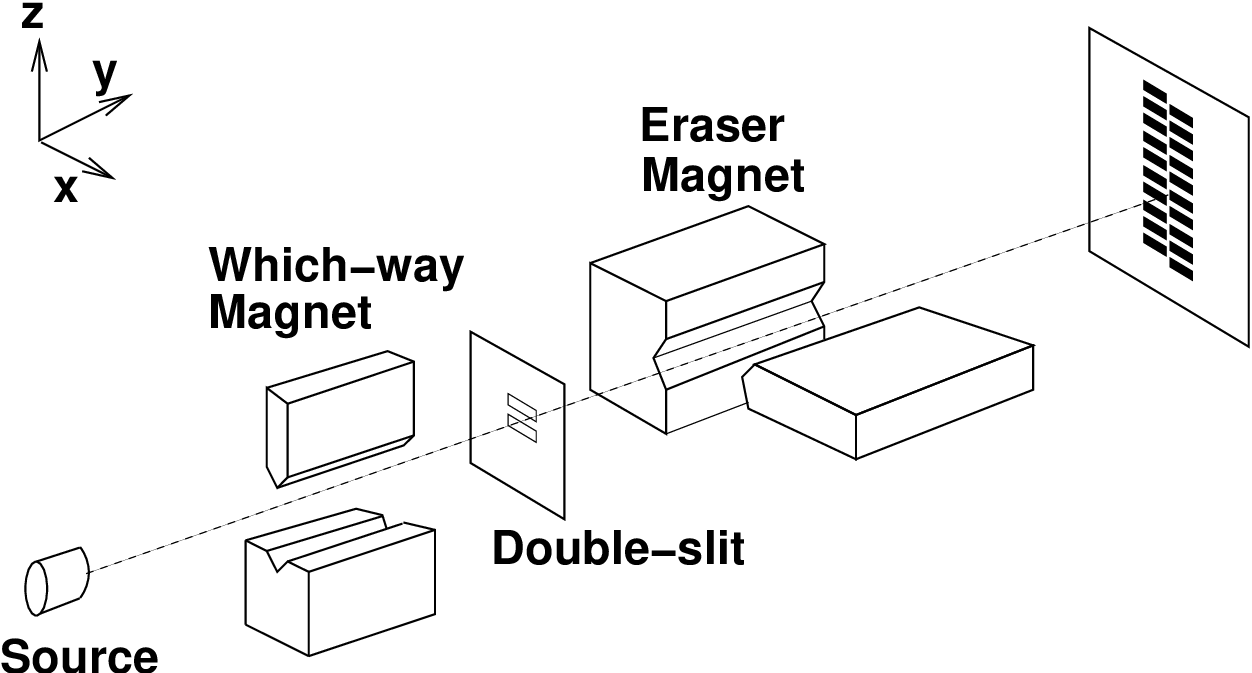}}}
\caption{ Schematic diagram of proposed quantum eraser. Which-way magnet
splits the beam into two so that they impinge on the double-slit. Eraser
magnet splits the interfering beams by pulling apart the eigenstates of
the x-component of the spin.}
\end{figure}

The setup consists of a Stern-Gerlach setup and a source of spin-1/2
particle (see Fig. 1). The particle travels along the positive y-axis,
and the magnetic field is along the z-direction. After the Stern-Gerlach magnet, which we will
call, the {\em which-way magnet}, there is a double-slit, kept such that the
slits are parallel to the x-axis.  Normally, in a 
double-split experiment, the position spread of the state of the
particle (along the direction of the slit) when it reaches the slit, should be larger than distance
between the two slits. Only then will both the slits ``see" the particle
at the same time. On the other hand, in the present setup, the initial
position spread of the particle state has to be much smaller than the distance
between  the slits. The particle starts out in a spin state 
${1\over\sqrt{2}}(|S_z;+\rangle+|S_z;-\rangle)$, where $|S_z;\pm\rangle$
represent the eigenstates of $\hat{S_z}$, the z-component of the spin.

The magnetic field of the which-way magnet entangles the position spatial
wave-function of the particle with the spin-states. Let us assume that
the state of the particle, when it reaches the double slit, is
\begin{equation}
|\Psi_i\rangle = {1\over\sqrt{2}}(|S_z;+\rangle|\phi_+\rangle
+ |S_z;-\rangle|\phi_-\rangle),
\end{equation}
where $|\phi_\pm\rangle$ correspond to spatial wave functions centered at
the upper and the lower slit, respectively.
When particle crosses the slits, which are much narrower than the position
spread of $|\phi_\pm\rangle$, the state which emerges on the other side of
the slits consists of wave-packets which are localized in a much narrower
region space. Consequently these packets spread much faster as the particle
travels in time. Beyond a certain distance after the double-slit, the
wave packets would have spread enough to overlap strongly with each other.
Suppose that the state of the particle, at this time, is given by
\begin{equation}
|\Psi_f\rangle = {1\over\sqrt{2}}(|S_z;+\rangle|\psi_+\rangle
+ |S_z;-\rangle|\psi_-\rangle),
\end{equation}
where $|\psi_\pm\rangle$ represent very spread out wave packets, which
strongly overlap with each other.
This is the region in which interference is expected, if which-way information
is not there. Although the wave-packets overlap with each other, each 
carries a which-way marker with it, in the form of the spin states
$|S_z;\pm\rangle$. Thus, if this particle is made to fall on a screen, no
interference will be seen. This can be verified by calculating the
probability of finding the particle at a position along the z-axis,
which should be
\begin{equation}
|\Psi_f(z)|^2 = {1\over 2}(|\psi_+(z)|^2 + |\psi_-(z)|^2).\label{nofringes}
\end{equation}
Of course, non-observation of interference by itself is no achievement,
as it can also happen
because of an imprecise experiment. Its underlying reason will be obvious only
if the interference pattern can emerge later, as we will see in the following.

Let us now introduce another Stern-Gerlach magnet, which we call the
{\em eraser magnet}, with field along the x-axis. The effect of the eraser
magnet will be to spatially separate out the components of the state
$|\Psi_f\rangle$ depending on spin eigenstates $|S_x;\pm\rangle$.
In order to analyze what will happen in such a situation, let us
write the state $|\Psi_f\rangle$ in terms of $|S_x;\pm\rangle$:
\begin{eqnarray}
|\Psi_f\rangle &=& {|S_x;+\rangle + |S_x;-\rangle\over 2}|\psi_+\rangle
+ {|S_x;+\rangle - |S_x;-\rangle\over 2}|\psi_-\rangle \nonumber\\
&=& |S_x;+\rangle{|\psi_+\rangle + |\psi_-\rangle\over 2}
   + |S_x;-\rangle{|\psi_+\rangle - |\psi_-\rangle\over 2} .
\end{eqnarray}
The eraser magnet will cause the piece of the wave function correlated
to $|S_x;+\rangle$ to shift along the positive x-axis, and that correlated
to $|S_x;-\rangle$ to shift towards the negative x-axis. When the particle
reaches the screen, the state acquires the form
\begin{equation}
|\Psi_e\rangle =  |S_x;+\rangle{|\psi_{1+}\rangle + |\psi_{1-}\rangle\over 2}
   + |S_x;-\rangle{|\psi_{2+}\rangle - |\psi_{2-}\rangle\over 2},
\end{equation}
where subscript $1$ indicates a wave-packet shifted towards the positive
x-direction, and the subscript $2$ indicates a wave-packet shifted towards
the negative x-direction. The strength of the eraser magnetic field
and the position of the screen are so adjusted that the wave-packets
with opposite shifts along the x-axis, have negligible overlap, i.e.,
$\langle\psi_{1\pm}|\psi_{2\pm}\rangle = 0$.

Let us calculate the probability of finding the particle at a point $(x,z)$
on the screen. Keeping in mind the orthogonality of $|S_x;\pm\rangle$,
this is given by
\begin{eqnarray}
|\Psi_e(x,z)|^2 &=&  {1\over 4}|\psi_{1+}(x,z)|^2 +  {1\over 4}|\psi_{1-}(x,z)|^2  \nonumber\\
            &+& {1\over 4}\psi^*_{1+}(x,z)\psi_{1-}(x,z)
            + {1\over 4}\psi^*_{1-}(x,z)\psi_{1+}(x,z)\nonumber\\
            &&+  {1\over 4}|\psi_{2+}(x,z)|^2 +  {1\over 4}|\psi_{2-}(x,z)|^2 \nonumber\\
            &-& {1\over 4}\psi^*_{2+}(x,z)\psi_{2-}(x,z)
            - {1\over 4}\psi^*_{2-}(x,z)\psi_{2+}(x,z)\nonumber\\
\label{erased}
\end{eqnarray}
So, how is the situation different from what it was before? If there were
no eraser magnet, $\psi_{1\pm}(x,z)$ would be the same as $\psi_{2\pm}(x,z)$,
and the cross terms in (\ref{erased}) would cancel out. In this situation
(\ref{erased}) would be identical to (\ref{nofringes}), which would mean,
no interference. But with the eraser
magnet on, the terms with subscripts $1$ and $2$ in (\ref{erased})
represent {\em two interference patterns} which are at different locations
on the x-axis! This is quantum erasure because particles reaching the
two different locations on the x-axis have spin-states $|S_x;+\rangle$
and $|S_x;-\rangle$ respectively, each of which possess no which-way
information by themselves.

The two interference patterns appear identical, but because the cross-terms
in one are with a negative sign, there would be a slight difference.
It can be demonstrated that the difference would be that one interference
pattern is vertically shifted with respect to the other, by one fringe.
This makes the two interference patterns complementary, in the sense that
the two combined result in no fringes, but only continuously varying intensity.
This is of course expected, and is also seen in other implementations of
quantum eraser.

\section{Example with Gaussian wave-packets}

If the analysis till now appears speculative, one can exemplify it by
a rigorous calculation. Let us start from the stage at which the particle
wave packets emerge from the double slit. We assume that the packets are
Gaussian. The state of the particle can be written down as
\begin{eqnarray}
\Psi(x,z) &=&  A e^{-{x^2\over 4\Omega^2}}\left(|S_z;+\rangle e^{-{(z-z_0)^2\over 4\sigma^2}}
+ |S_z;-\rangle e^{-{(z+z_0)^2\over 4\sigma^2}}\right)\nonumber\\
&&
\end{eqnarray}
where $A = {1\over\sqrt{4\pi\sigma\Omega}}$. It represents two Gaussians
centered at $z = \pm z_0$, where $2z_0$ is the
distance between the two slits. The Gaussians centered at $z = \pm z_0$ are
entangled with the spin states $|S_z;\pm\rangle$ respectively.

Suppose that the momentum in the $y$ direction is such that the particle
takes a time $t$ to reach the screen. During this evolution, the packets
would have spread. We let the state evolve under the influence of a free
Hamiltonian $p_x^2/2m + p_z^2/2m$.  The state of the particle, at a time
$t$ is given by
\begin{eqnarray}
\Psi(x,z,t) &=&  A_t
 e^{-{x^2\over 4(\Omega^2-{it\hbar\over 2m})}}
\left(|S_z;+\rangle e^{-{(z-z_0)^2\over 4(\sigma^2-{it\hbar\over 2m})}}\right.
\left.+ |S_z;-\rangle e^{-{(z+z_0)^2\over 4(\sigma^2-{it\hbar\over 2m})}}\right),
\end{eqnarray}
where $A_t = {(4\pi)^{-1/2}\over\sqrt{(\sigma-{it\hbar\over 2m\sigma})
(\Omega-{it\hbar\over m\Omega})}}$.
The probability density of the particle hitting the point (x,z) on the screen
is given by
\begin{eqnarray}
|\Psi(x,z,t)|^2 &=&  |A_t|^2
e^{-{x^2\over 2\Omega_t^2}}
\left(e^{-{(z-z_0)^2\over 2\sigma_t^2}}\right.
\left.+ e^{-{(z+z_0)^2\over 2\sigma_t^2}}\right) .
\end{eqnarray}
where $\Omega_t^2 = \Omega^2+{t^2\hbar^2\over m^2\Omega^2}$ and
$\sigma_t^2 = \sigma^2+{t^2\hbar^2\over m^2\sigma^2}$.
This represents no interference of particles, which is expected because 
each particle carries with itself a which-way marker in the form of spin
state $S_z;\pm\rangle$ (see Fig. 2(a)).

In order to see what happens when the eraser magnet is switched on, we
write the state of the particle in terms of the eigenstates of $\hat{S}_x$,
$S_x;\pm\rangle$:
\begin{eqnarray}
|\Psi_f\rangle &=& {1\over\sqrt{2}}|S_x;+\rangle
A e^{-{x^2\over 4\Omega^2}}\left(e^{-{(z-z_0)^2\over 4\sigma^2}}
+ e^{-{(z+z_0)^2\over 4\sigma^2}}\right)\nonumber\\
 &+& {1\over\sqrt{2}}|S_x;-\rangle A e^{-{x^2\over 4\Omega^2}}
\left(e^{-{(z-z_0)^2\over 4\sigma^2}}
- e^{-{(z+z_0)^2\over 4\sigma^2}}\right)
\end{eqnarray}
The particle starts out from the double-slit at time $t=0$ and enters 
the region of the eraser magnet at $t=t_i$ and leaves it at time $t=t_i+t_e$.
Then it travels and reaches the screen at a time $t$. Outside the region
of the eraser magnet, the Hamiltonian governing the particle is given by
$\hat{H}_f = \hat{p}_x^2/2m + \hat{p}_z^2/2m$. Within the region of the
eraser magnet, the particle experiences an inhomogeneous magnetic field in
the x-direction.  The Hamiltonian in this region is given by
$\hat{H}_e = \hat{p}_x^2/2m - \beta x\sigma_x + \hat{p}_z^2/2m$. The time
evolution under the influence of this Hamiltonian can be worked out
explicitly \cite{av}. The
state of the particle, when it reaches the screen is given by \cite{av}
\begin{eqnarray}
\Psi_e(x,z,t) &=& 
{1\over\sqrt{2}}A_t |S_x;+\rangle e^{{-{(x-{\beta\over 2m}t_e^2)^2\over 4(\Omega^2-{it\hbar\over 2m})}}
+{i\beta t_ex\over\hbar}} 
\left(e^{{-{(z-z_0)^2\over 4(\sigma^2-{it\hbar\over 2m})}}}
+ e^{-{(z+z_0)^2\over 4(\sigma^2-{it\hbar\over 2m})}}\right) \nonumber\\
&&+ {1\over\sqrt{2}}A_t |S_x;-\rangle e^{{-{(x+{\beta\over 2m}t_e^2)^2\over 4(\Omega^2-{it\hbar\over 2m})}}
-{i\beta t_ex\over\hbar}} 
 \left(e^{{-{(z-z_0)^2\over 4(\sigma^2-{it\hbar\over 2m})}}}
- e^{-{(z+z_0)^2\over 4(\sigma^2-{it\hbar\over 2m})}}\right) ,
\end{eqnarray}
The above expression has a simple
interpretation. The first term represents a Gaussian centered at
$x = {1\over 2}{\beta\over m}t_e^2$, which is just the distance traveled
by a particle in time $t_e$ with an acceleration $\beta/m$. The third term
represents a Gaussian centered at $x = -{1\over 2}{\beta\over m}t_e^2$,
which is the distance traveled
by a particle in time $t_e$ with an acceleration $-\beta/m$. The terms
$\exp(\pm {i\over\hbar}\beta t_e x)$ indicate that momentum of the particle
is $\pm\beta t_e$ which should be the momentum acquired by a particle after
being accelerated for a time $t_e$ with an acceleration $\pm\beta/m$.
The probability density of the particle hitting the screen at a point
$(x, z)$ is given by
\begin{eqnarray}
|\Psi_e(x,z,t)|^2&=& 
{|A_t|^2\over 2} e^{-{(x-{\beta\over 2m}t_e^2)^2\over 2\Omega_t^2}}
[P_+(z) + P_-(z) + 2 f(z) ] \nonumber\\
&& + {|A_t|^2\over 2} e^{-{(x+{\beta\over 2m}t_e^2)^2\over 2\Omega_t^2}}
[P_+(z) + P_-(z) - 2 f(z) ] 
\end{eqnarray}
where
\begin{equation}
P_\pm(z) = \exp\left[{-{(z\mp z_0)^2\over
2(\sigma^2+{t^2\hbar^2\over m^2\sigma^2})}}\right],~~~
f(z)= \exp\left[{-{(z^2+z_0^2)\over
2(\sigma^2+{t^2\hbar^2\over m^2\sigma^2})}}\right]
\cos\left[{2z z_0 t\hbar/m\sigma^2 \over
\sigma^2+{t^2\hbar^2\over m^2\sigma^2}}\right].
\end{equation}
We plot this distribution and find that indeed the interference pattern,
which was lost because of the which-way information, now appears because the
eraser magnet has erased the which-way information (see Fig. 2(b)). Notice
that the z-position of the dark fringe of one pattern is the same as a
bright fringe of the other, so that if they were not x-shifted, they would
have cancelled out.

\begin{figure}
\centerline{\resizebox{7.0cm}{!}{\includegraphics{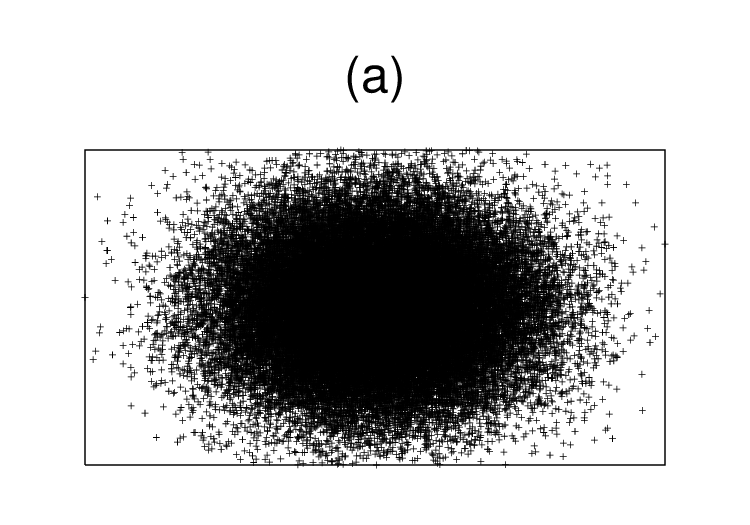}}
\resizebox{7.0cm}{!}{\includegraphics{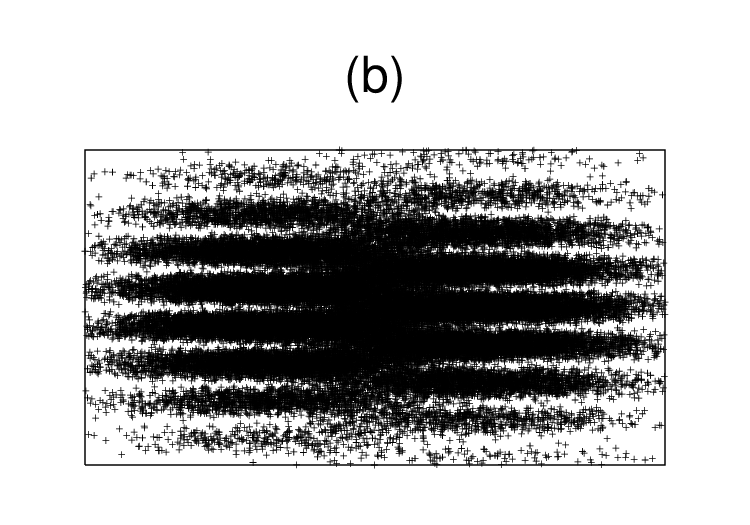}}}
\caption{ Probability density distribution of particles hitting the screen
(a) when the eraser magnet is switched off, and 
(b) when the eraser magnet is switched on.  }
\end{figure}

A few points need to be made at this stage. Without the eraser magnet, even
though the which-way information is carried by the particle, one might wonder
if
one can actually extract it. The answer is that it can be done by
putting a magnet in place of the eraser magnet, which has a magnetic field
pointing along the z-axis, {\em but inhomogeneous along the x-axis}. For
example $\vec{B} = \hat{k} B_0 x$. This would cause the particles in states
$|S_z;+\rangle$ and $|S_z;-\rangle$ to separate out along the x-axis.

It may be instructive to compare this proposal with other implementations of
quantum eraser. In most other
implementations, one doesn't get the interference directly and has to
do a coincident counting of particles with certain states of the which-way
detectors. Some people have this feeling, that the interference pattern is
actually lost for good and one is only picking it out from the erased pattern
in an artificial way. In this respect, this method has the advantage that
one can observe the interference appear right before one's eyes as the
eraser magnet is switched on. Another point is that 
this method allows one to demonstrate quantum erasure using
massive particles, instead of photons, using a simple setup. 

In conclusion, we have proposed a new implementation of quantum eraser
using spin-1/2 particles in a modified Stern-Gerlach setup. The which-way
information can be erased simply by applying a magnetic field, and two
complementary interference patterns appear on the screen. The method
doesn't require
any fancy setup or entangled sources, except a Stern-Gerlach apparatus.

\end{document}